# MARS15 SIMULATION OF RADIATION ENVIRONMENT AT THE ESS LINAC *†

**N.V. Mokhov[1#], Yu.I. Eidelman[2], I.L. Rakhno[1], L. Tchelidze[3], I.S. Tropin[1]**

[1]Fermi National Accelerator Laboratory, Batavia IL 60510-5011, USA

[2]Euclid Techlabs LLC, Solon OH 44136-1866, USA

[3]European Spallation Source, Tunavagen 24, 223 63 Lund, Sweden

## ABSTRACT

*Comprehensive studies with the MARS15(2016) Monte-Carlo code are described on evaluation of prompt and residual radiation levels induced by nominal and accidental beam losses in the 5-MW, 2-GeV* European Spallation Source (*ESS) Linac. These are to provide a basis for radiation shielding design verification through the accelerator complex. The calculation model is based on the latest engineering design and includes a sophisticated algorithm for particle tracking in the machine RF cavities as well as a well-established model of the beam loss. Substantial efforts were put in solving the deep-penetration problem for the thick shielding around the tunnel with numerous complex penetrations. It allowed us to study in detail not only the prompt dose, but also component and air activation, radiation loads on the soil outside the tunnel, and skyshine studies for the complicated 3-D surface above the machine. Among the other things, the newest features in MARS15 (2016), such as a ROOT-based beamline builder and a TENDL-based event generator for nuclear interactions below 100 MeV, were very useful in this challenging application.*

---

*Work supported by Fermi Research Alliance, LLC under contract No. DE-AC02-07CH11359 with the U.S. Department of Energy.

†Presented paper at the 13th Meeting of the task-force on Shielding aspects of Accelerators, Targets and Irradiation Facilities (SATIF-13), HZDR, October 10-12, 2016, Dresden, Germany

#mokhov@fnal.gov

# MARS15 SIMULATION OF RADIATION ENVIRONMENT AT THE ESS LINAC

**Nikolai V. Mokhov[1], Yury I. Eidelman[2], Igor L. Rakhno[1], Lali Tchelidze[3], Igor S. Tropin[1]**
[1]Fermi National Accelerator Laboratory, Batavia, Illinois 60510, USA
[2]Euclid Techlabs LLC, Solon OH 44136-1866, USA
[3]European Spallation Source, Tunavagen 24, 223 63 Lund, Sweden

**Introduction**

The European Spallation Source (ESS) accelerator complex is under construction in Lund, Sweden. It will deliver a 5-MW, 2-GeV proton beam for the ESS facility. The first ESS shielding calculations were performed in 2013-2014 [1]. The purpose of the current work was to build the calculation model of the entire machine as close to the reality as possible – with all the information on geometry, materials and beamline element electromagnetic fields taken into account - and perform detailed simulations to verify that as built ESS accelerator tunnel building together with the planned soil structure around provides sufficient attenuation for the normal operational prompt radiation. The Linac tunnel walls and soil together are referred to as a shielding and its adequacy is checked for normal operations. Weak spots, such as personnel or equipment penetrations, were carefully modelled and studied. Duration of an accidental point beam loss for which the shielding would prevent excessive exposure to workers outside of it was determined as well.

The existing geometry model of the ESS Linac was updated with the newest detailed information – based on the engineering design - on beam-line components such as quadrupole magnets, accelerating cavities and cryomodules as well as as-built Linac tunnel, numerous associated systems, and shielding layout. The study described in this paper was performed with the newest version of the MARS15(2016) code [2,3]. The new model uses the ROOT-geometry [4] mode in MARS15 and includes all the details available on geometry, materials and electromagnetic fields as well as the standard model of the beam loss [5]. The powerful features in MARS15 (2016), such as a new ROOT-based beamline builder, a TENDL-based event generator for nuclear interactions below 100 MeV, and variance reduction techiques were especially useful in this challenging application.

The paper starts with a description of the ESS machine MARS model. Then, beam loss assumptions for the normal operation are outlined. This is followed by a section on the particle acceleration model developed and its verification. The results of the study for the normal operation are presented on the prompt dose in the tunnel and associated systems as well as on machine component and air radio-activation. A special attention is given to the skyshine radiation calculation on the ESS site. Assumptions and results for the worst-case beam accident conclude the paper.

**MARS15(2016) model**

Included in the MARS15 model is the newest information – based on the engineering design - on beam-line components such as quadrupole magnets, accelerating cavities and cryomodules as well as as-built Linac tunnel, associated systems and shielding layout with all the details available on geometry, materials and electromagnetic fields. This information from the ESS CAD teams was obtained in the form of STEP files and converted to the MARS15 ROOT geometry modules. Location of beam line components is determined from the optics files of the linac lattice. The element sequence was built and linked to the above geometry modules and to materials/field maps by means of the advanced ROOT-based Beamline Builder. The following penetrations in the shielding were included in the model:

- 27 stubs for RF waveguides and cables (including waveguides, cables and filler material for additional shielding (in some cases))
- 4 personnel emergency exits
- 1 cryogenic transfer line
- 1 HEBT loading bay
- 1 A2T-GSA chicane
- 1 HVAC exhaust duct
- 4 alignment penetrations

ESS-specific descriptions of 32 materials are used in the model. Various fragments of the MARS15 calculation geometry model are given in Figures 1-3.

**Figure 1: A side view of the ESS accelerator and adjacent structures (left) and details of the berm in a lateral view (right) in the MARS15 model**

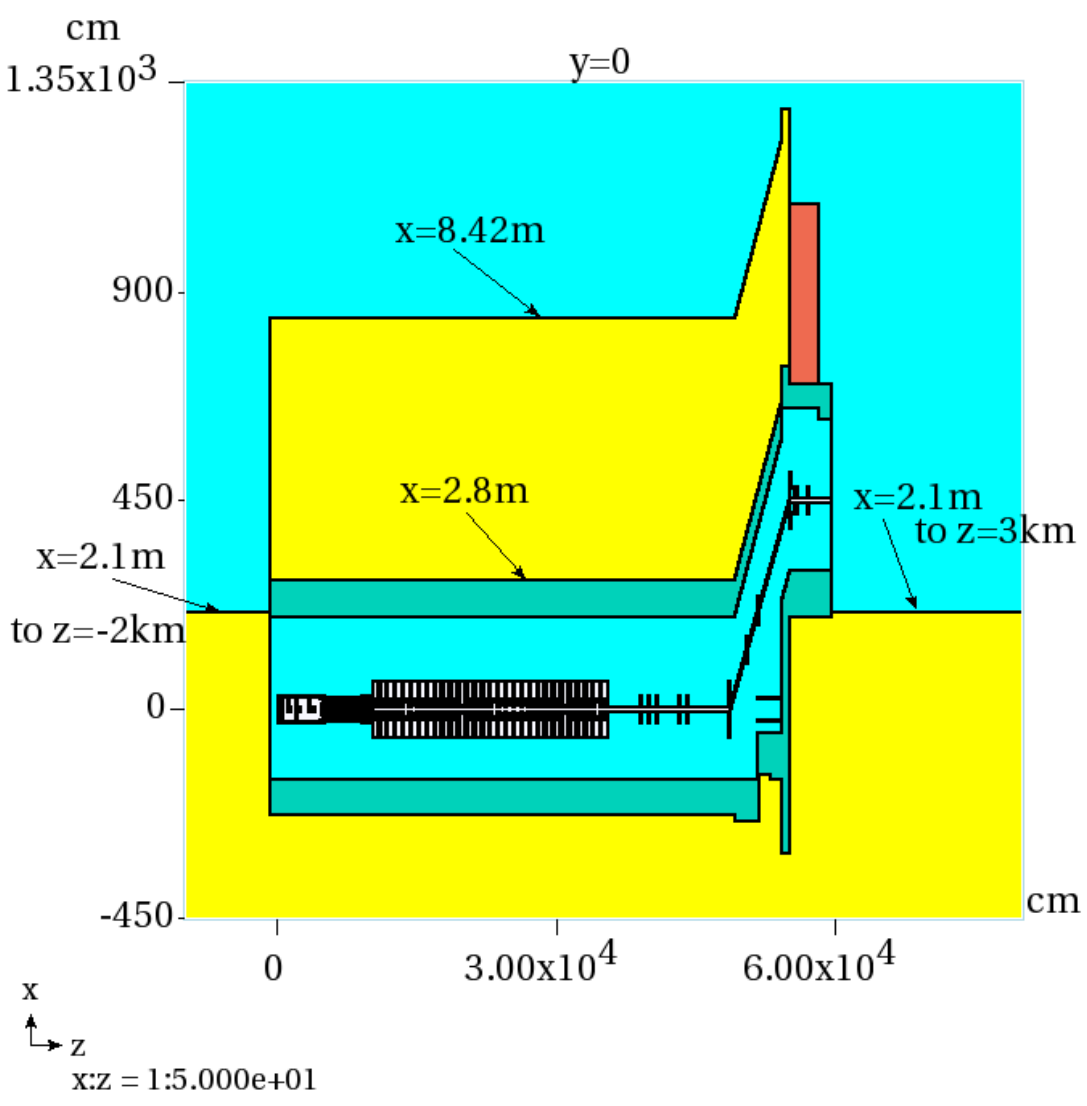

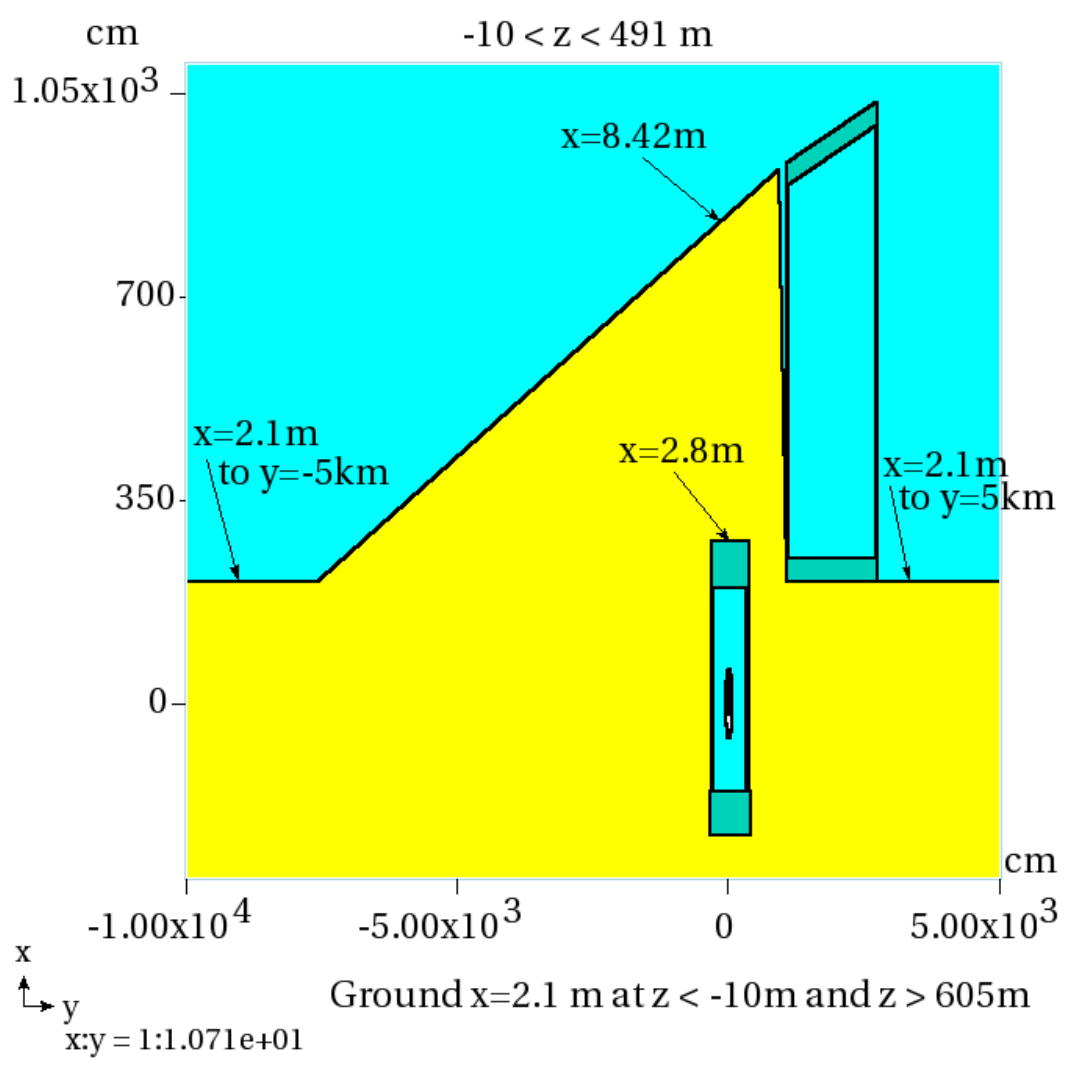

**Figure 2: A plan view of the ESS accelerator and adjacent structures (left) and implementation of a typical stub and filling scheme (right)**

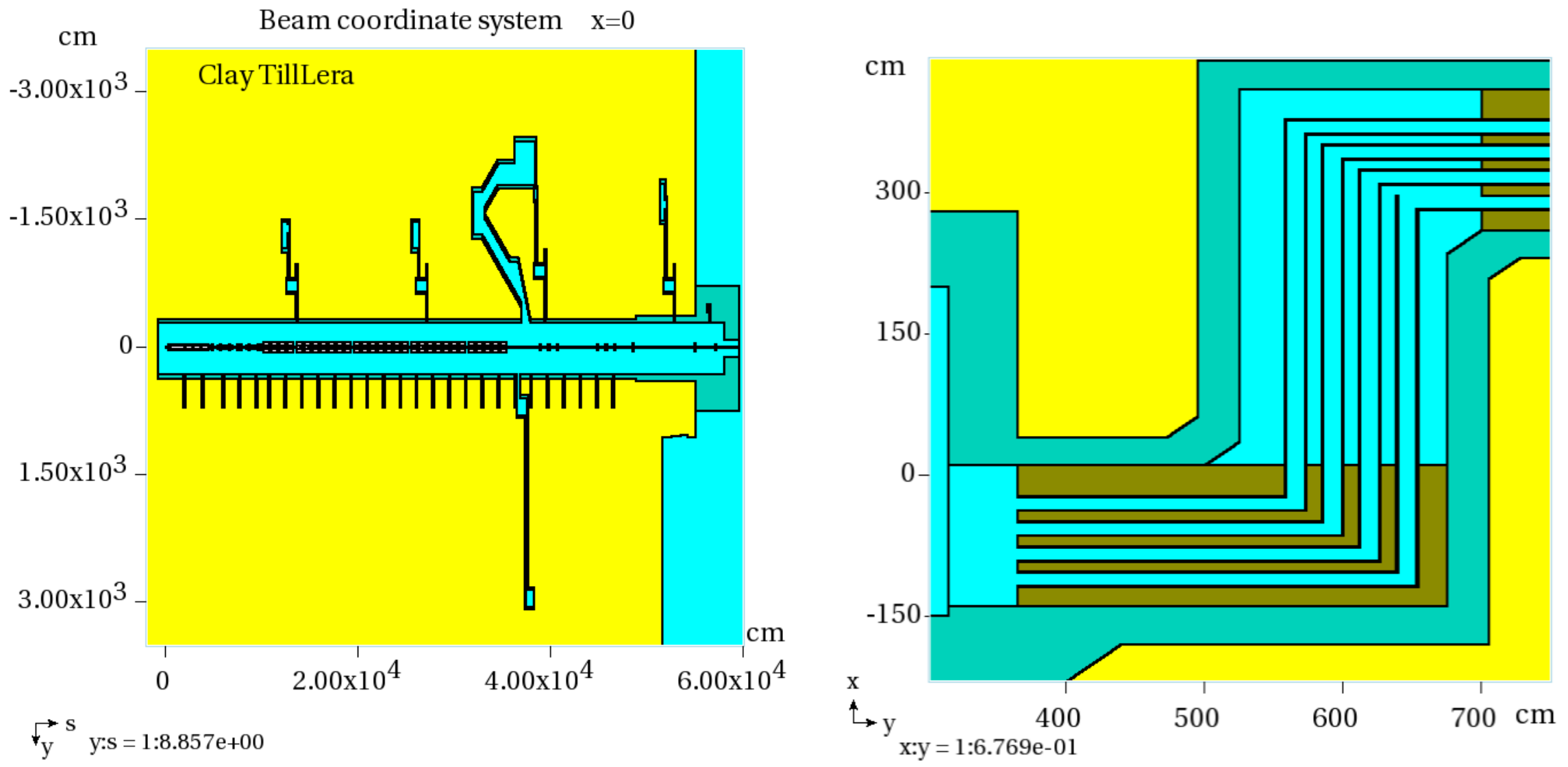

**Figure 3: 3D representation of a high-beta cavity(left) and quadrupole geometry and magnetic field distributions in MARS15 model (right)**

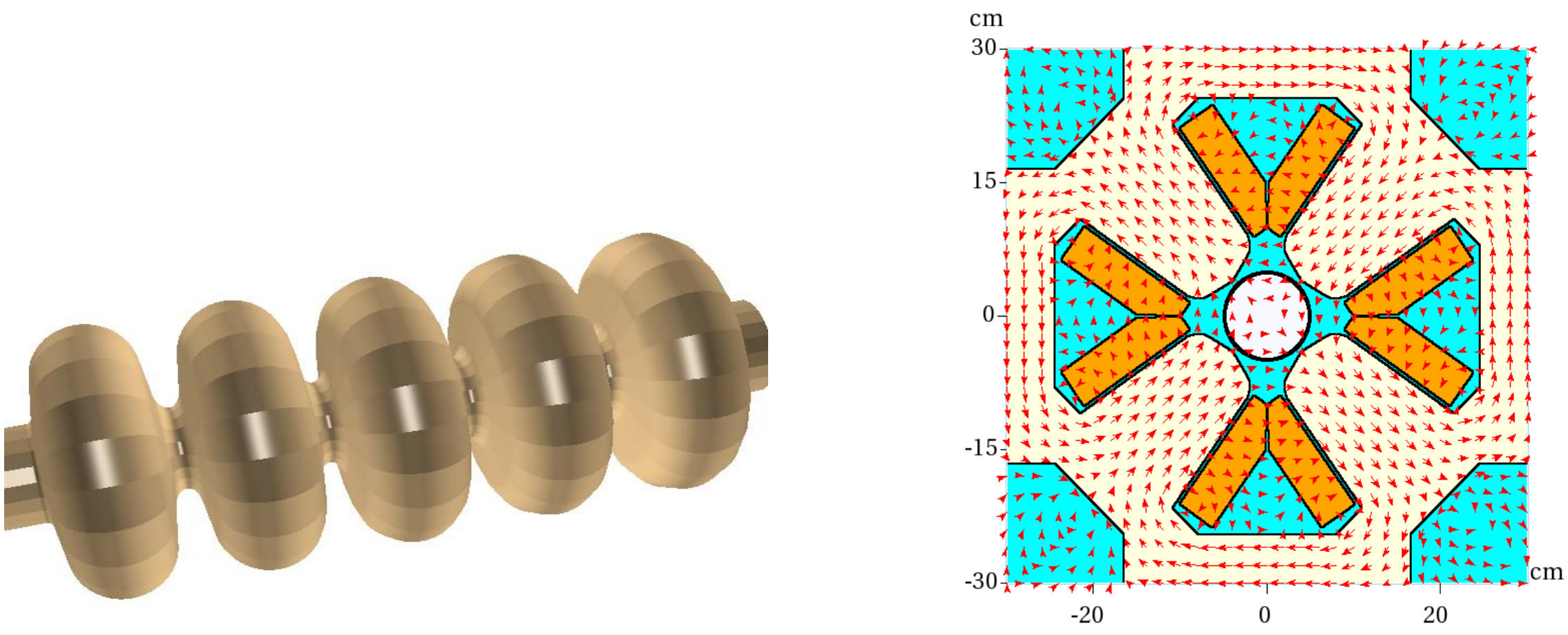

Customized steppers with the high-order Runge-Kutta algorithms were created for optimal particle tracking in the drift tube linac (DTL), spoke cavities, medium- and high-beta elliptical Superconducting Radio-Frequency (SRF) cavities, quadrupole/dipole magnets and thick shielding of a quite complex 3D configuration. For inelastic nuclear interactions, a combination of the LAQGSM and CEM event generators [6] were used at 0.15 – 2 GeV, extended TENDL [7] based generators below 0.15 GeV down to 1 MeV for charged particles and 14 MeV for neutrons, and ENDFB-VI libraries [8] for neutrons below 14 MeV. Energy thresholds were 0.001 eV for neutrons and 100 keV for charged particles and photons. Two variance reduction techniques were used in the shielding simulations: multi-stepping and splitting/Russian roulette.

To calculate the effective prompt dose in the regions of interest from the particle/energy dependent track-length estimator in the course of Monte-Carlo simulations, a standard MARS15 set of flux-to-dose (FTD) conversion factors was used. That is the data from ICRP103 [9] supplied with ICRP60 ones [10] with data for neutrons from [11] and for other particles from [12]. A proposed at ESS set of FTD factors [13] was found to be very similar for the components of radiation field outside the Linac shielding. The set [13] was used in this study for the skyshine dose calculations.

## Beam loss model

A proton beam loss rate of 1 W/m is assumed to be distributed uniformly longitudinally and azimuthally on the beampipe at a 3-mrad grazing angle in the beam reference system. It was considered as a design criteria and an upper limit during normal operations. This was derived from the hands-on maintenance conditions [5] as adopted at ESS. Beam energy dependence on location along the ESS accelerator is shown in Figure 4 along with the modelled 1 W/m beam loss rate.

**Figure 4: Primary proton beam energy as a function of a location along the linac (left) and modelled normal operational 1 W/m beam loss rate (right)**

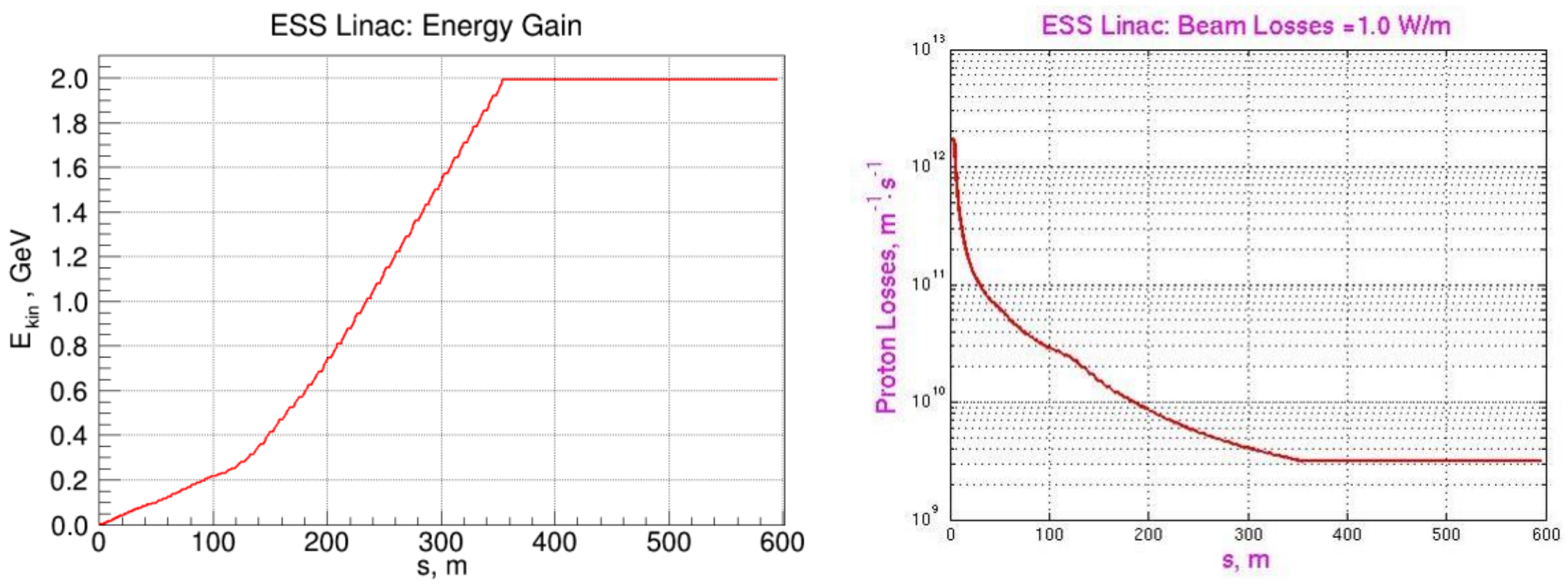

## Modelling acceleration

In this study, a "Kick and Deflection" Model (KDM) is based on calculating the change in kinetic energy of the particles when they are passing through a single cell of the cavity (kick) and at the same time they are deflected in the moment of impact. The deflection is determined by the conservation of the transverse momentum of the particles upon impact while their total momentum is changed due to the change in energy. Figure 5 (left) shows the electrical field inside a single cell. In the linac modelling, it is taken into account that the phase and amplitude for each cavity are different along the lattice. Figure 5(right) shows the amplitude distributions in the spoke, medium-beta and high-beta sections of the linac.

**Figure 5: E-field inside the cell (left) and amplitude distribution along the linac (right)**

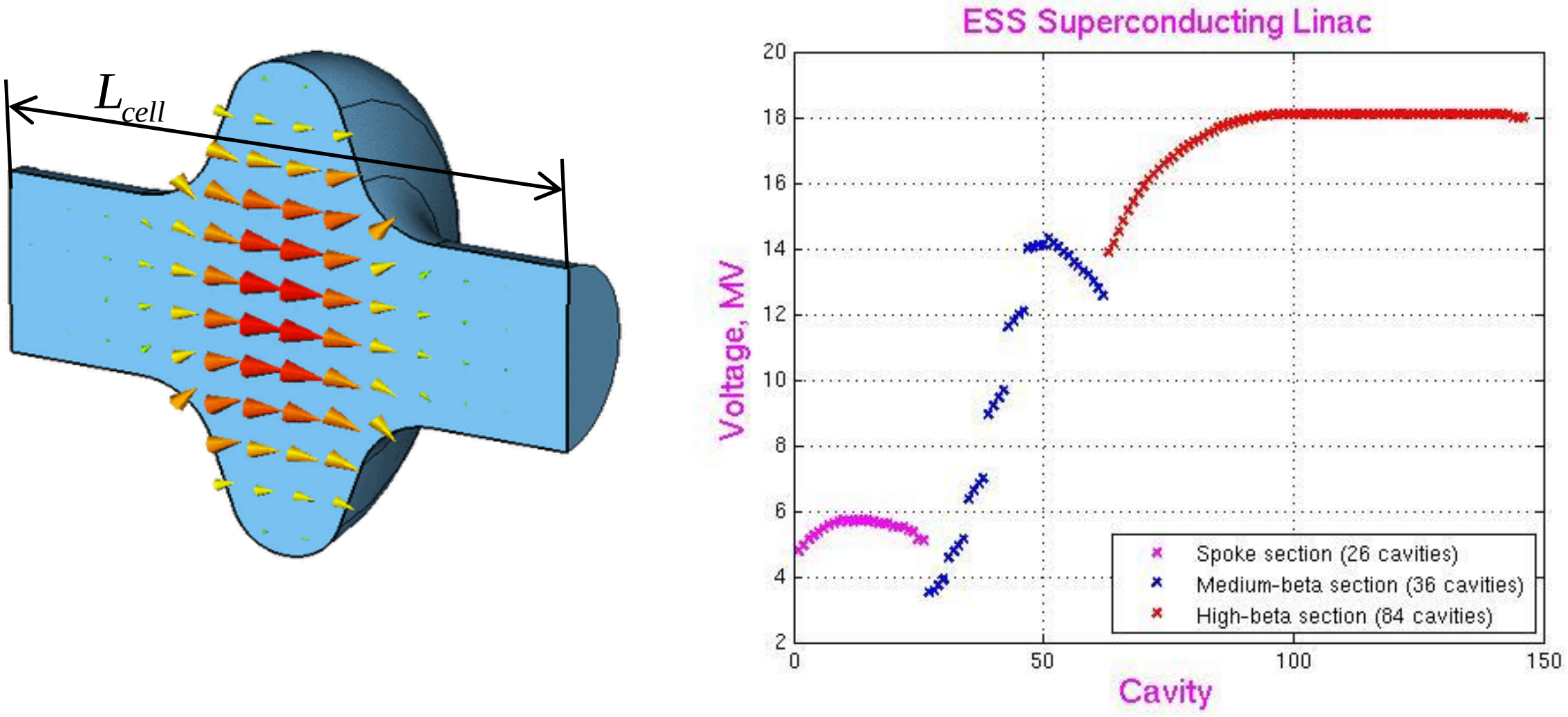

The ASTRA code (A Space charge TRacking Algorithm [14]) was used to verify the acceleration model employed in this study. The comparison of the particle acceleration and their trajectories in the cavities of all types was made. Figure 6 illustrates a very good agreement between the two schemes.

**Figure 6: MARS and ASTRA comparison of acceleration in the first high-beta module (left) and tracks inside the first medium-beta module (right)**

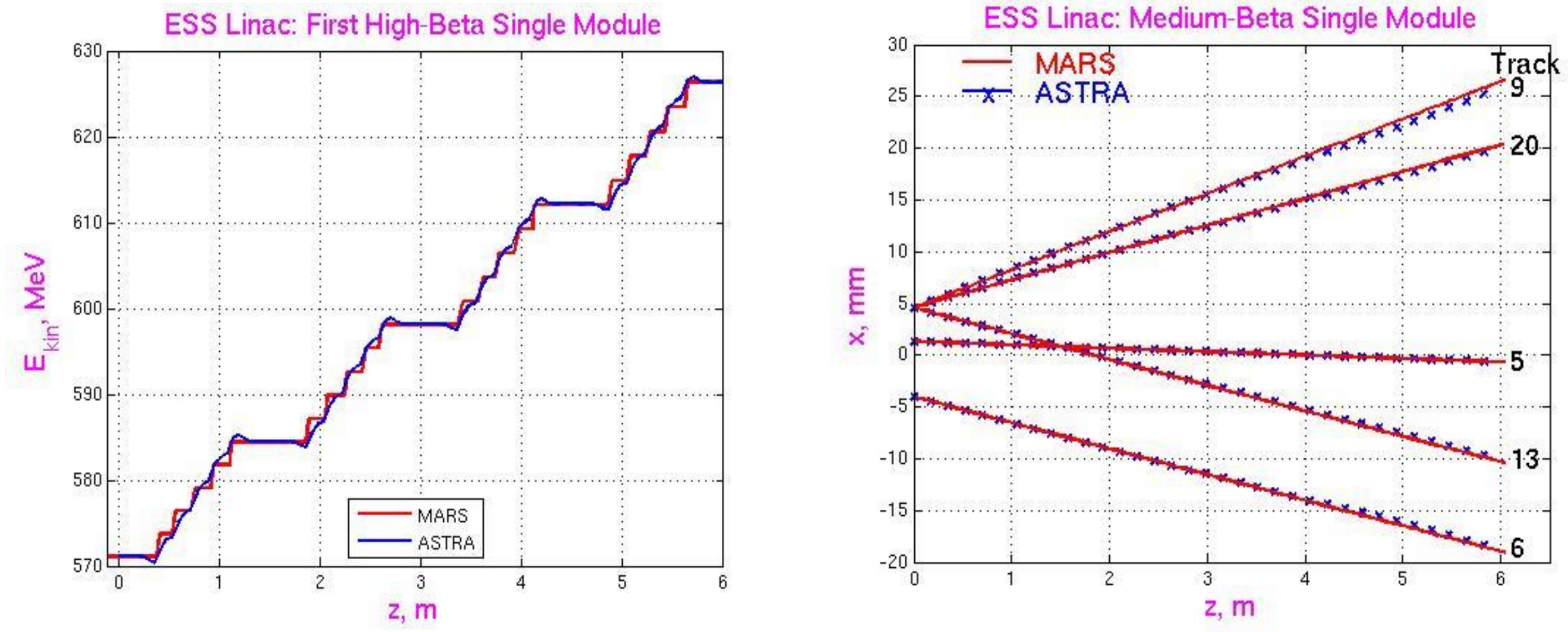

## Operational scenario: beam loss of 1 W/m

### *Prompt dose in the linac and its associated systems*

The calculated prompt dose distributions in the linac tunnel, associated systems, surrounding soil and above the berm are presented here. Note that the color palette is the same in all isocontour plots of this section. To get statistically valid results on prompt dose on the berm above the ESS machine and create a source term for skyshine, the MARS15 runs included scoring source terms at the outside of the tunnel concrete walls. Those source terms are represented by files with detailed information on particles entering the soil around the ESS linac tunnel walls. These are used then in the second stage of MARS15 simulations as a source. The properties of such sources along with a comprehensive set of results on prompt dose at the ESS complex are described in detail in [16]. Energy of particles entering the soil can be as high as 0.8-0.9 GeV for photons and electrons and 1.5 GeV for protons and neutrons. At the same time, the mean energies are substantially lower: 18.6 MeV (n), 163 MeV (p), 13.6 MeV (photons) and 51 MeV (electrons).

Figure 7 shows total and neutron only dose rate isocontours in a vertical plane through the beam axis. The DTL section is the least hot region of the accelerator. The ratio of the highest and lowest prompt dose rate in the immediate vicinity of the beamline (at the Linac end and beginning of DTL, respectively) is approximately 100. The maximum dose rate on the berm is between 0.1 and 1 µSv/h for the majority of the ESS linac. It reaches 1 µSv/h at the end of the linac straight horizontal tunnel section as can be seen in Figure 8 that shows the prompt dose rate at four longitudinal locations of the straight horizontal tunnel as a function of elevation from the beam in soil above the tunnel concrete ceiling (at x = 280 cm) with the berm/air boundary of x = 840 cm at y=0. The dose stays at the level of 1 µSv/h through the end of the A2Tdogleg berm. A neutron contribution to the total dose on the berm is at a 97% level, while in the tunnel it is about 10%.

**Figure 7: Total (left) and neutron only (right) prompt dose rate isocontours in the entire ESS machine: vertical scan through the beamline center**

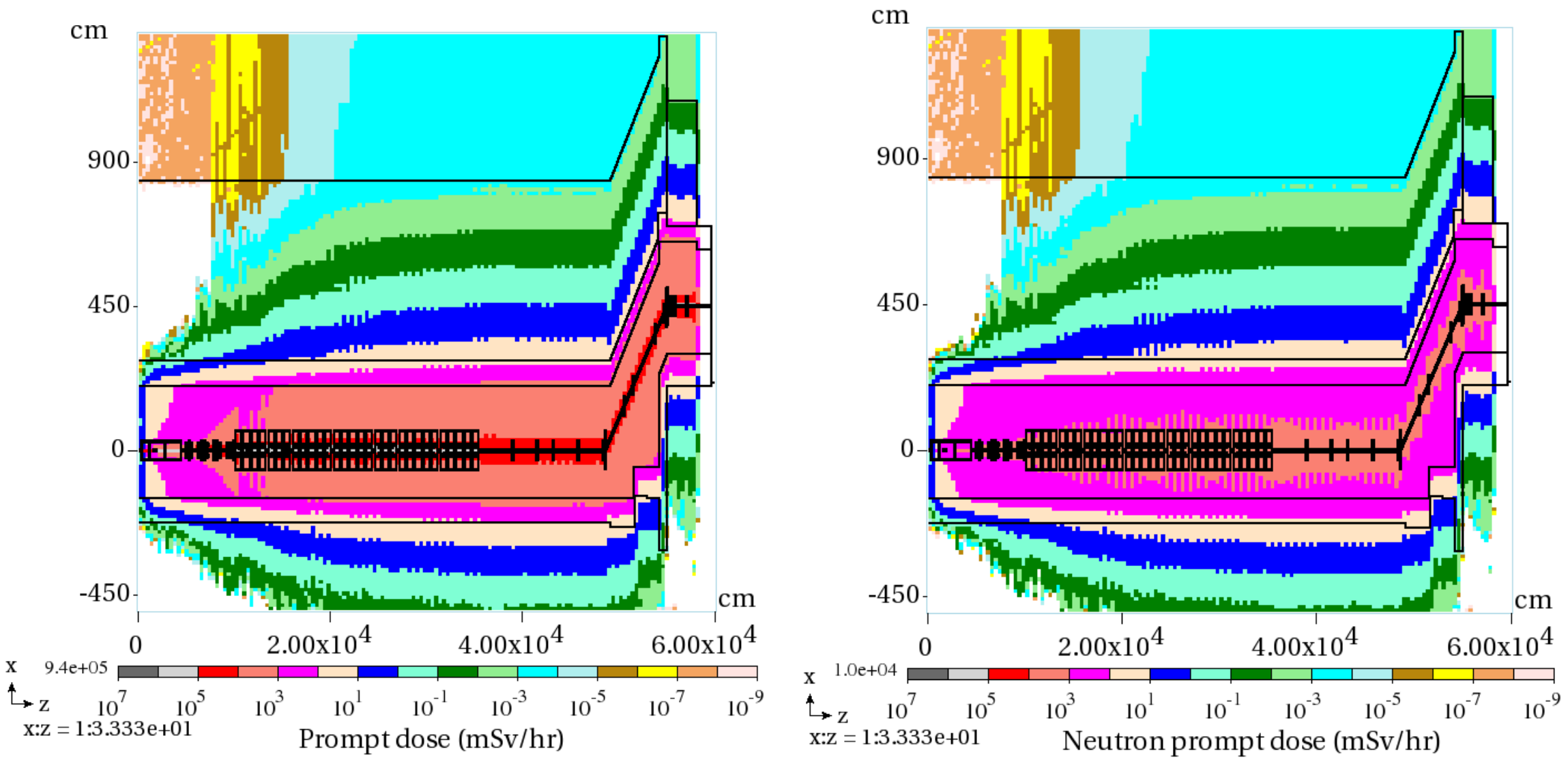

**Figure 8: Prompt dose rate at four longitudinal locations of the straight horizontal tunnel as a function of elevation in soil above the tunnel concrete ceiling (at x = 280 cm)**

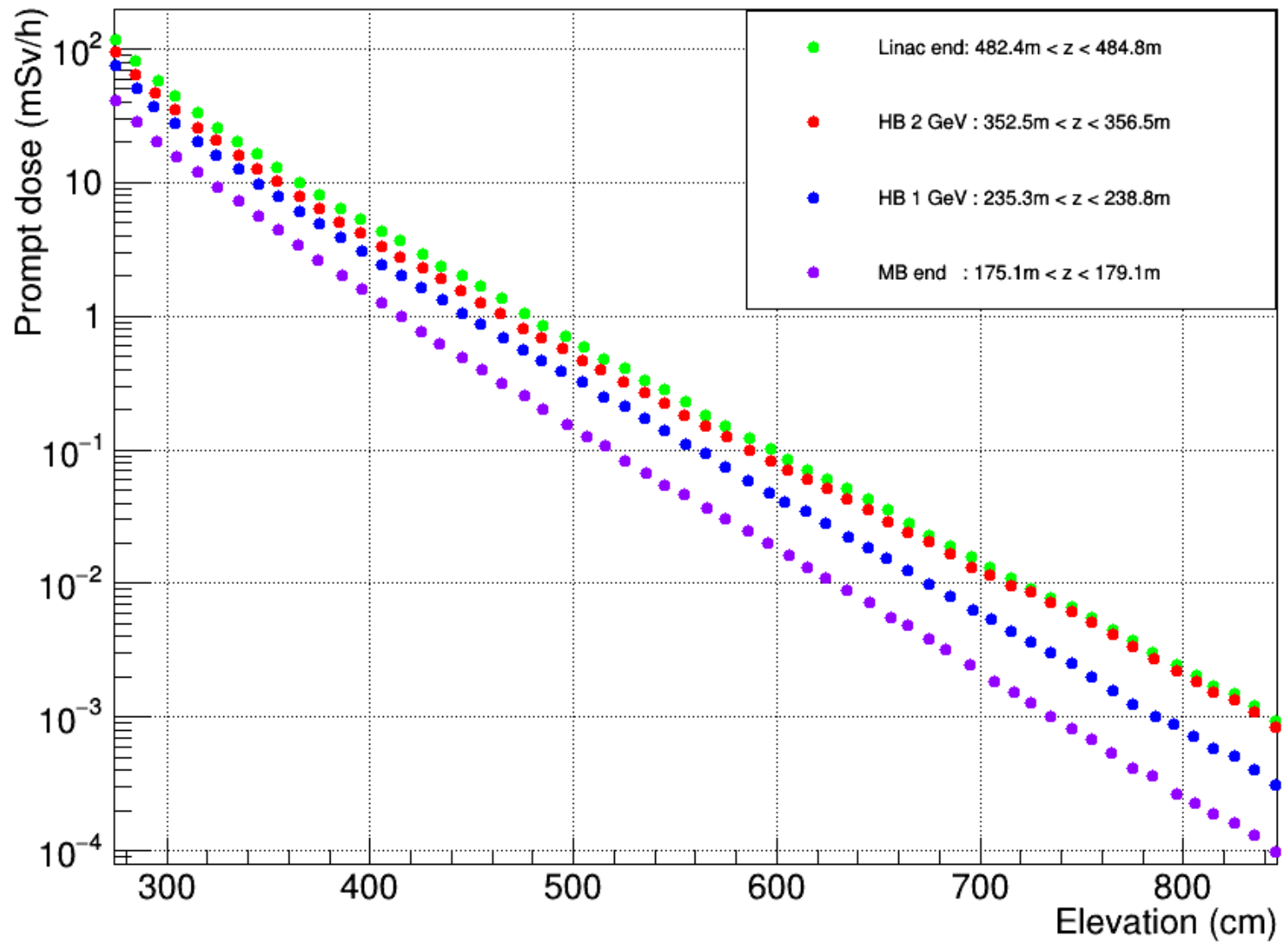

The transverse total dose distributions are shown in Figure 9 for two lngitudinal locations in the linac where the proton beam reaches 1 GeV in the high-beta section at z ~ 237 m (left) and at the end of the straight horizontal tunnel, z=483.5 m, E=2 GeV (right).

**Figure 9: Transverse scans of total prompt dose rate at a 1-GeV location in the high-beta section (left) and at 2-GeV in the end of the straight horizontal tunnel (right)**

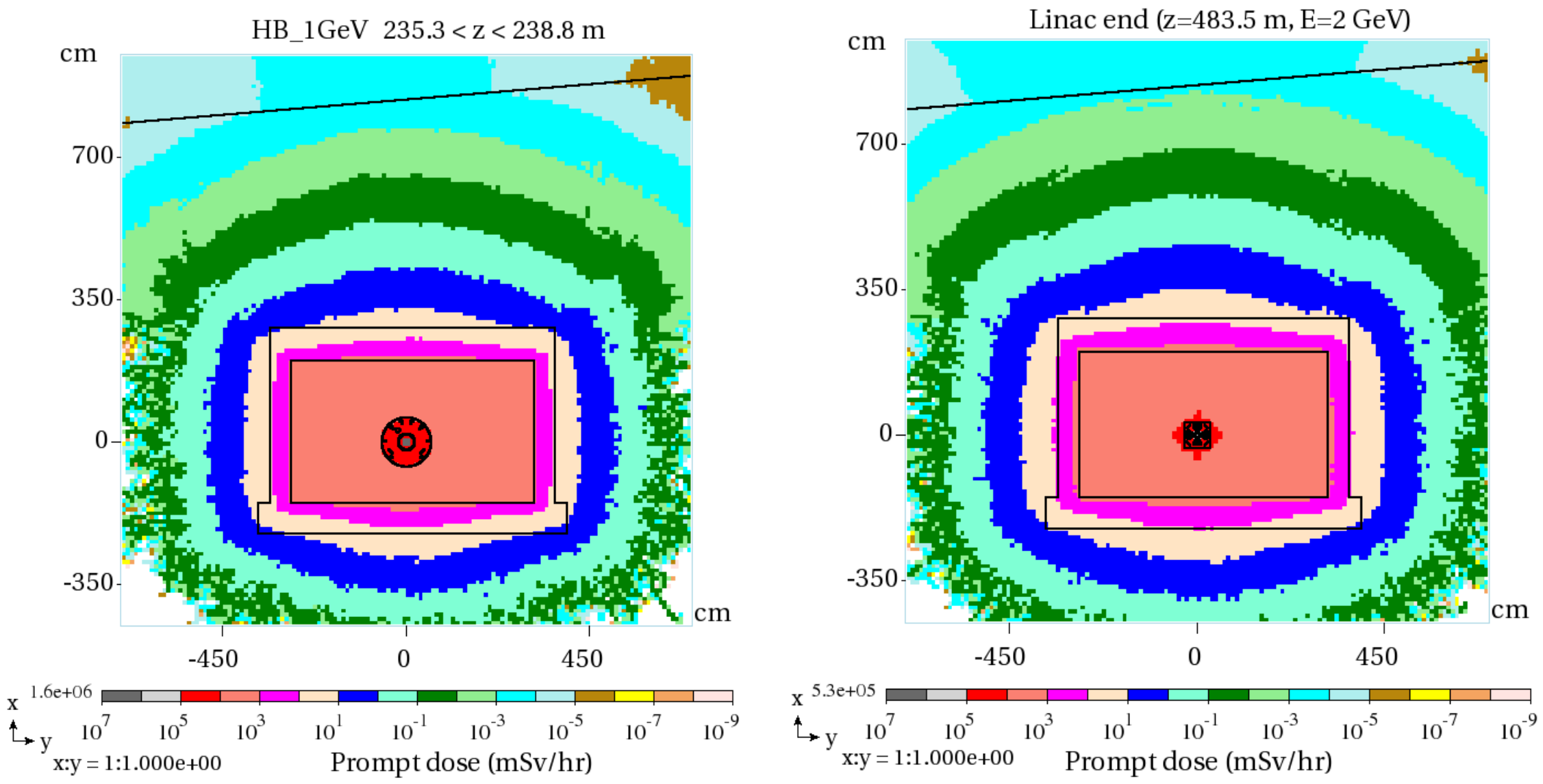

The prompt dose distributions for the normal operation were also calculated in klystron gallery and stubs (with and w/o filler), at emergency exits, in HEBT loading bay, A2T-GSA penetration, HVAC exhaust pipes, alignment penetrations, and cryogenic transfer line [15]. It was justified that with the current design and corresponding protective measures – such as sand/polybead filler in the stubs – the dose at the critical locations is kept below the target value of 1 μSv/hr in all the cases. As an example, Figure 10 shows the dose rate calculated at seven locations along the HEBT loading bay. The data refers to dose rates averaged over several penetration cross sections, and the distance from the beamline refers to horizontal distance between centre of each cross section and the beamline. The symbols in the geometry fragment indicate approximate locations of the centres of the cross sections used for the dose averaging.

**Figure 10: HEBT loading bay geometry fragment (left) and prompt dose in its section (right).**

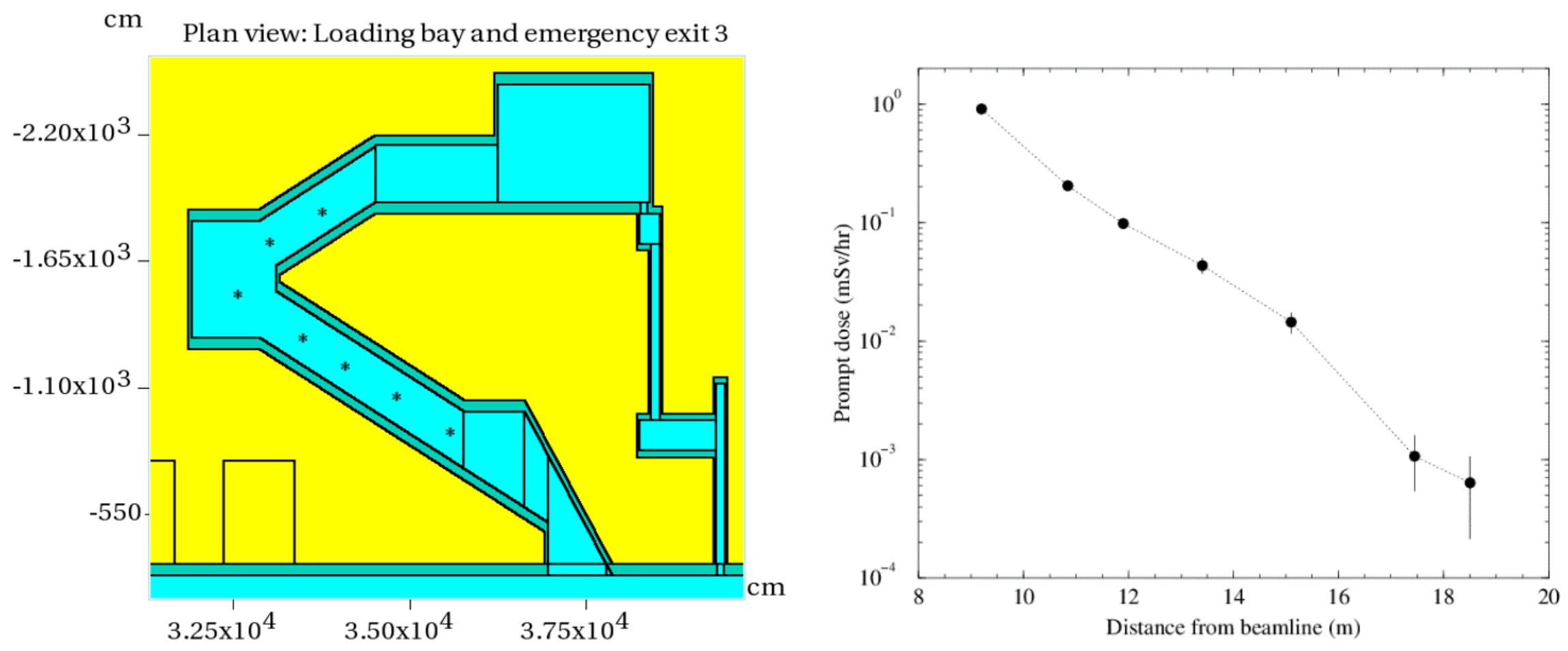

### *Component activation*

Residual dose rate on the outer surface of magnets – after 100 days of irradiation and 4 hours of cooldown – ranges from 0.3 to 1 mSv/h (see Figure 11), in a full consistency with the 1 W/m rule [5]. At the same conditions, the maximum contact dose on the concrete walls is 0.01 mSv/h. Air activation in the tunnel is described in [16].

**Figure 11: Residual dose rate on contact through the hottest cross-sections of quadrupole at the end of the high-beta section (left) and the dipole magnet at the beginning of the dogleg (right)**

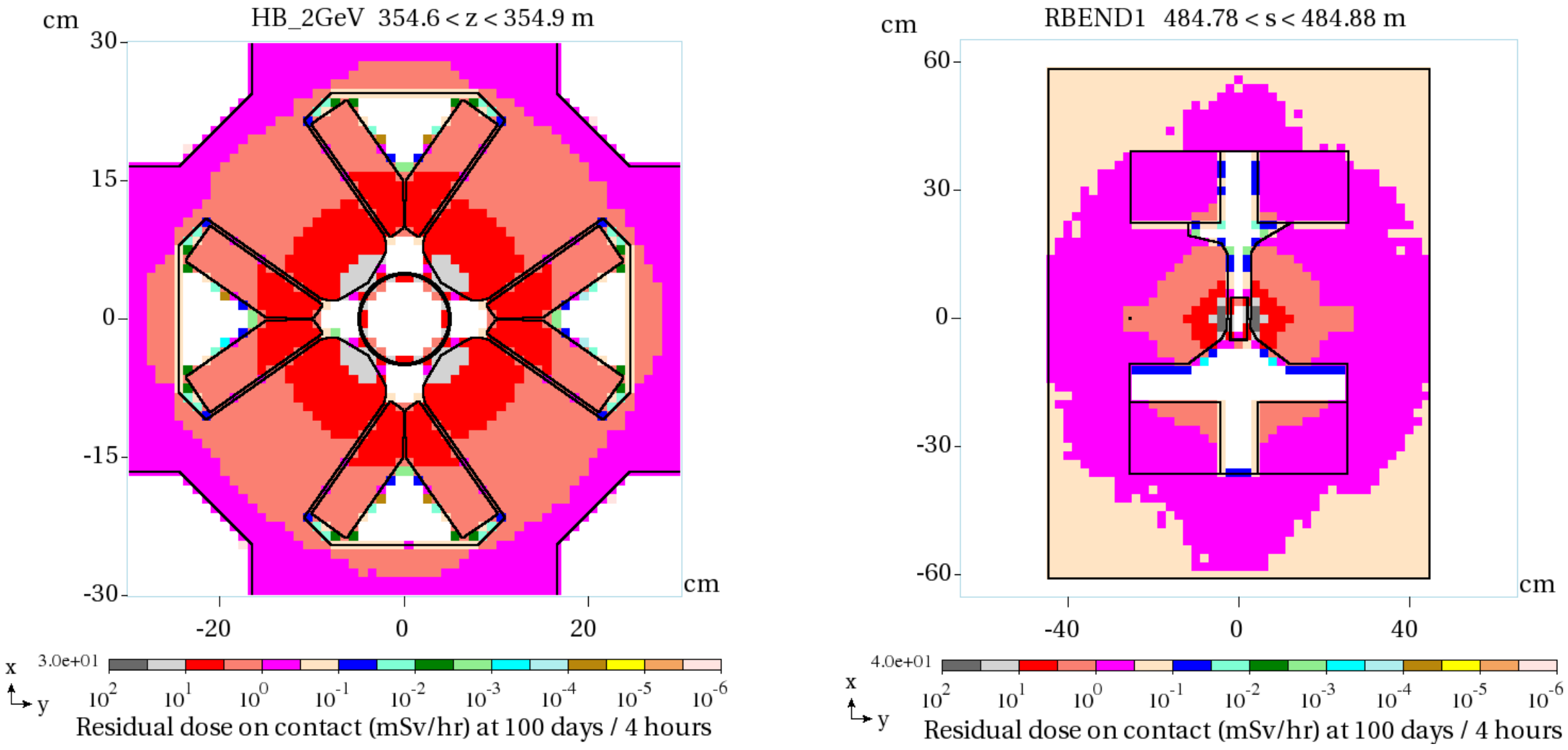

***Skyshine radiation***

Reference [17] describes a comprehensive study of skyshine dose maps at and around the ESS site associated with the ESS accelerator normal operations. Only a contribution from the beam losses in the linac (from zero to the end of the accelerator to the target region, up to neutron shield wall) was considered there. The primary particles beyond this point (neutron shield wall) were not tracked.

The source for skyshine calculations was tallied at the second stage of the deep penetration and source term MARS15 calculations as discussed in detail in Ref. [15, 16]. The coordinate system origin (x=y=z=0) is at the beginning of the ESS linac. A soil berm above the accelerator tunnel has a noticeable right (positive y) to left (negative y) slope, with the highest vertical coordinate being x=9.3 m. As seen from Figure 1, the berm height is at x=8.42m above the beam (y=0), with the overall ground level at the site at x=2.1 m. The skyshine calculations are done in the atmosphere with the air density gradually decreased with height according to the NASA Earth Atmosphere Model [18]. In this model, the air volume is centered at the source and is 10 km in height and 5 km in radius. In this study, the HVAC exhaust chicane inside the A2T region was filled with ordinary concrete of density 2.35 gcc. Thus, a contribution to skyshine from the HVAC exhaust penetration is not calculated and will require a further investigation. Skyshine from the target system is considered elsewhere and not included in the current report. The skyshine calculation is done with the MARS15(2016) code for particles exiting the soil berm, concrete structures and penetrations and entering the atmosphere air. These particles are scored in a separate file. It comprises parameters of neutrons, protons and photons with energies down to 0.001 eV (neutrons) and 100 keV (protons and photons). It was found in [15] - and was confirmed in the skyshine study - that the total dose rate on the ESS linac berm is 97% due to neutron-induced reactions and remaining 3% - due to photons with practically no contribution from other particles. Results below are obtained with neutron and photon flux-to-dose conversion factors of [13].

The prompt dose maps in the 1.7-m thick (human height) air layer right above the ESS linac berm is shown in Figure 12 (left). The dose maps are superimposed with the contours of the linac systems taken at x=8 m above the beam. The peak dose right above the berm at y=0 is below 0.01 μSv/hr in the DTL, Spoke and the first half of medium-beta (z<150 m), 0.01 to 0.1 μSv/hr in the rest of medium-beta and beginning of high-beta (150<z<220 m), 0.1 to 1 μSv/hr in the rest of the Linac and up to the end of the A2T dogleg beamline (220<z<550 m). The peak dose above the berm in the latter region rapidly drops by an order of magnitude left and right of the upward beam projection. The small box in magenta corresponds to a section of HEBT loading bay (see Figure 13), inside of which, the radiation dose rates are attenuated by the concrete structure. Figure 12 (right) shows the skyshine dose map in the same central region just a few meters higher. Compare it to the distributions given in Figure 14 for the larger region -500 < y < 500 m, -100 < z < 1100 m, covering most of the ESS site. Figure 15 shows elevation and transverse skyshine radiation profiles. Much more results of this ESS skyshine studies can be found in Ref. [17].

**Figure 12: Prompt dose rate distributions in the 9.3 < x < 11 m (left) and 13.8 < x < 20 m (right) air layers above the ESS linac central region superimposed with the contours of the linac systems taken at 8 meters**

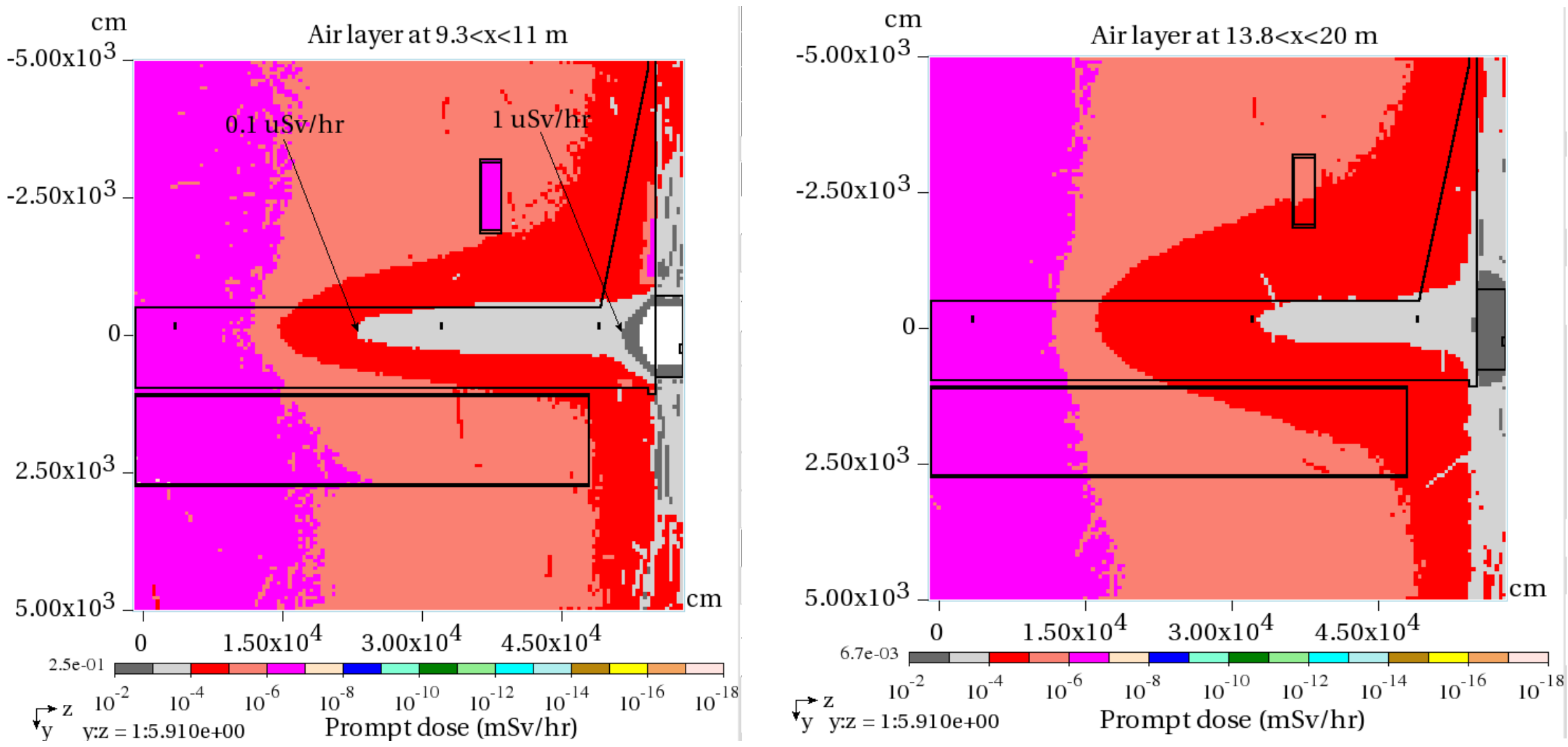

**Figure 13: A cross-sectional xy-view of the HEBT loading bay at z=370 m**

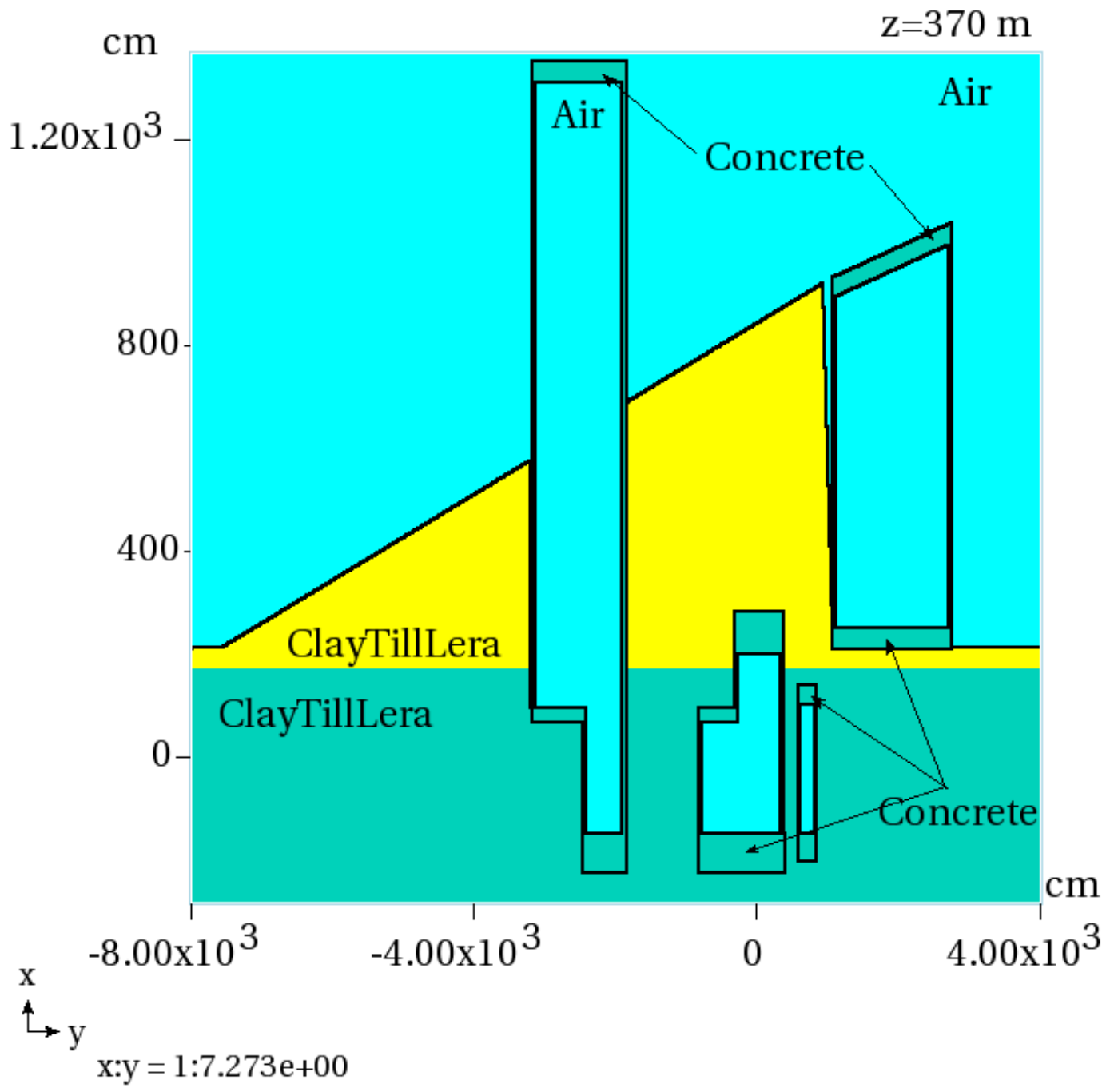

**Figure 14: Prompt dose rate distribution in the 13.8 < x < 20 m (left) and 20 < x < 100 m air layers above the ESS site superimposed with the contours of the linac systems taken at 8 meters**

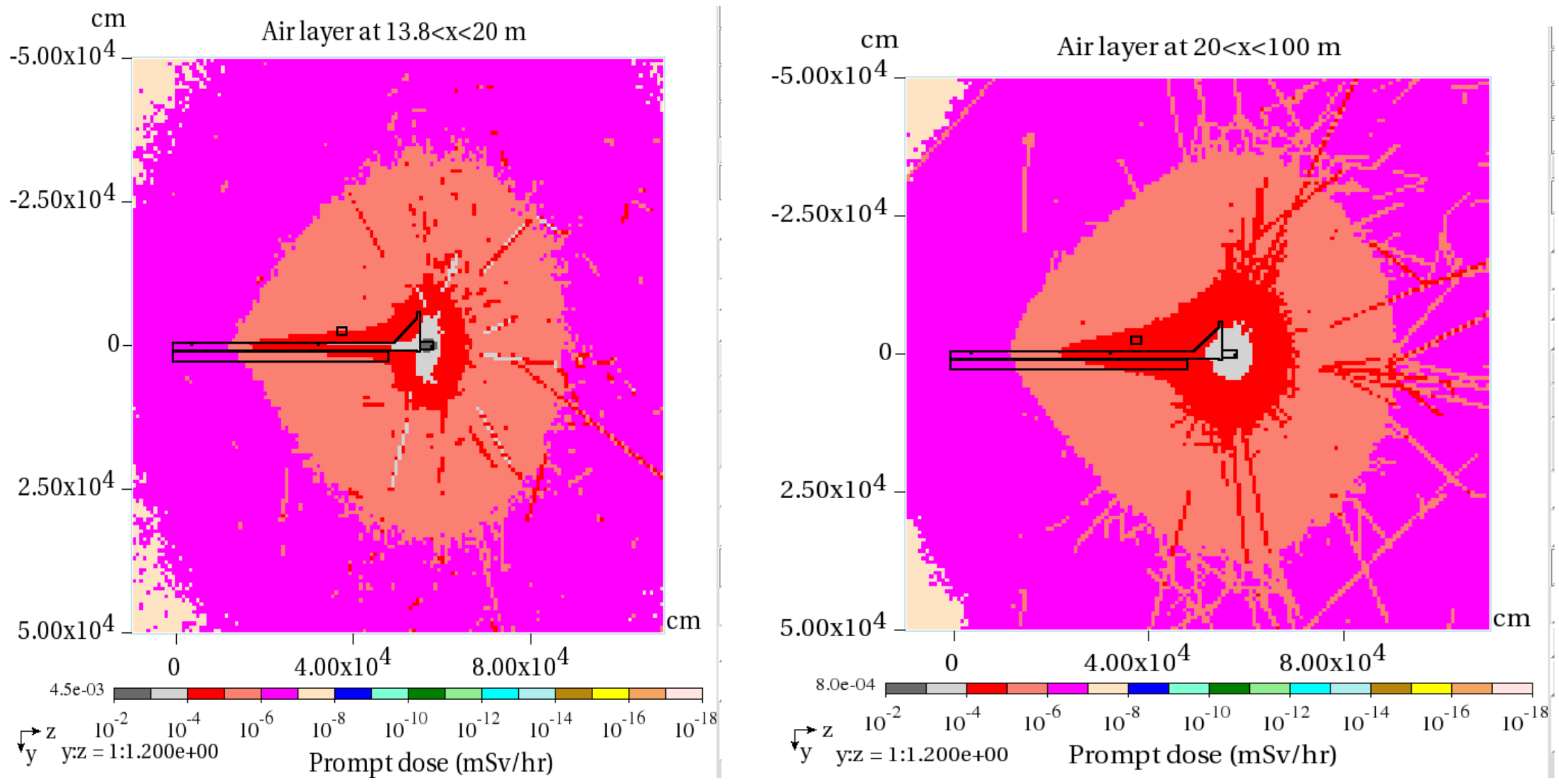

**Figure 15: Prompt dose rate xz-isocontours in the first kilometer above the Linac at -2.5 < y < 2.5 m (left) and xy-isocontours at the end of the linac 550 < z < 580 m (right).**

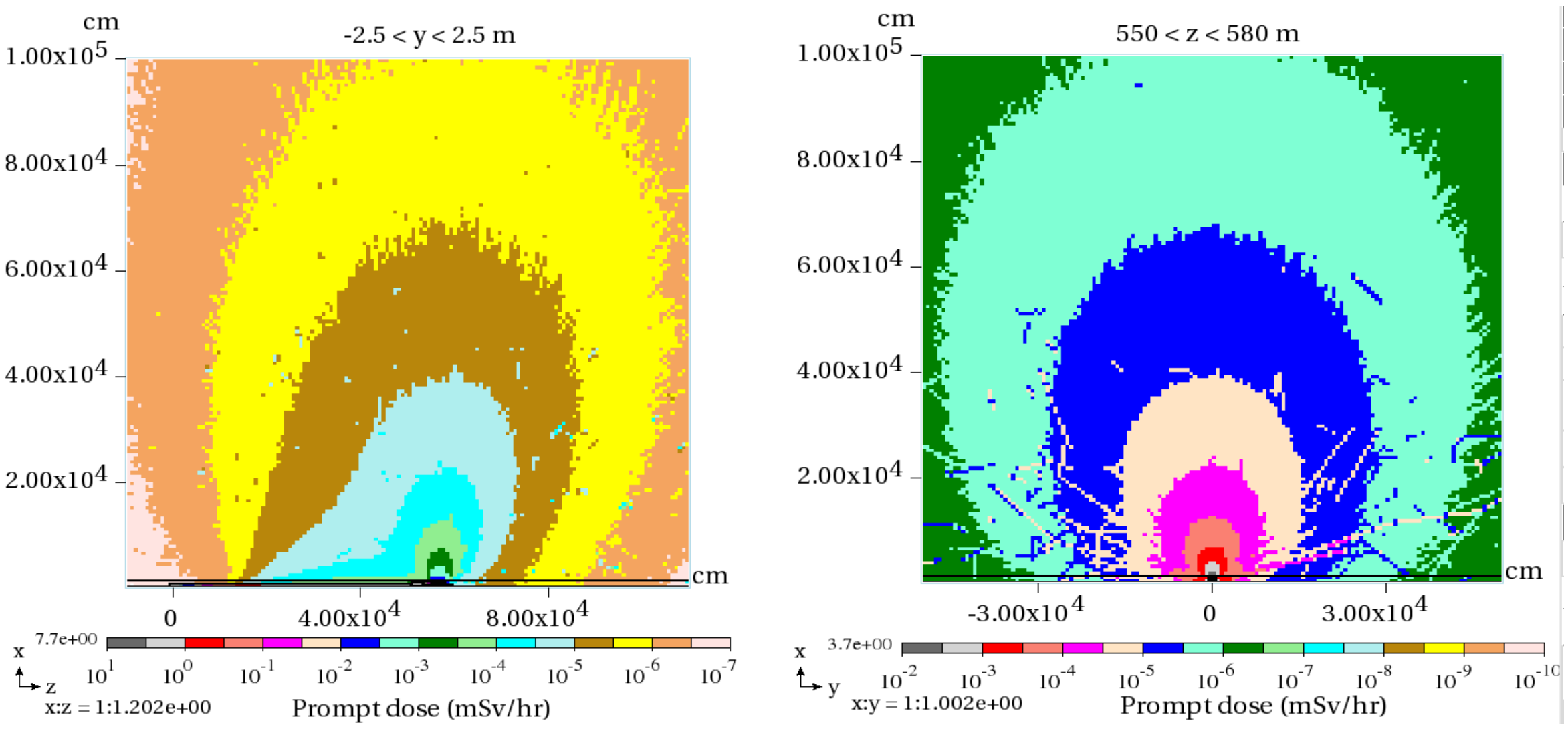

## Beam accident

The consideration was given to the impact on radiation environment of the accidental point loss of the full beam [15]. The primary concern is the corresponding peak prompt dose rate on top of the berm and in the hottest point in klystron gallery. A preliminary MARS15 analysis has shown that the worst case corresponds to a 2 GeV beam lost downstream of the high-beta linac at z ~ 447 meters just upstream of the quadrupole magnet quad-475 in front of the stub mouth. It is assumed that the point-like beam hits the beam pipe at a grazing angle of 3 mrad upward. There is a 2-fold increase in the dose rate for the worst stub if the loss point is in a horizontal plane towards the stub. The rate of 2 GeV protons that corresponds to the full 5 MW beam is $1.5625\times10^{16}$ p/s.

The prompt dose rate distribution for such a full beam loss event is shown in Figure 16 (left). The peak dose rate on the berm of 200 mSv/h was derived from similar histograms and the dose attenuation law obtained with results used for Figure 8. The prompt dose rate distribution in the lattice components and at the entrance to the stub at z = 448 m is shown in Figure 16 (right). With sand+poly filling in the first and third legs of the stub at this location, the peak dose at the entrance to the klystron gallery is 20 mSv/h for the full beam accident. The peak power density in the coils of the hottest quadrupole at this location is about 40 W/cm$^3$.

**Figure 16: Elevation view of prompt dose rate isocontours in the ESS linac tunnel and surrounding soil (left) and plan view of the dose at the entrance to the stub at z=448 m (right)**

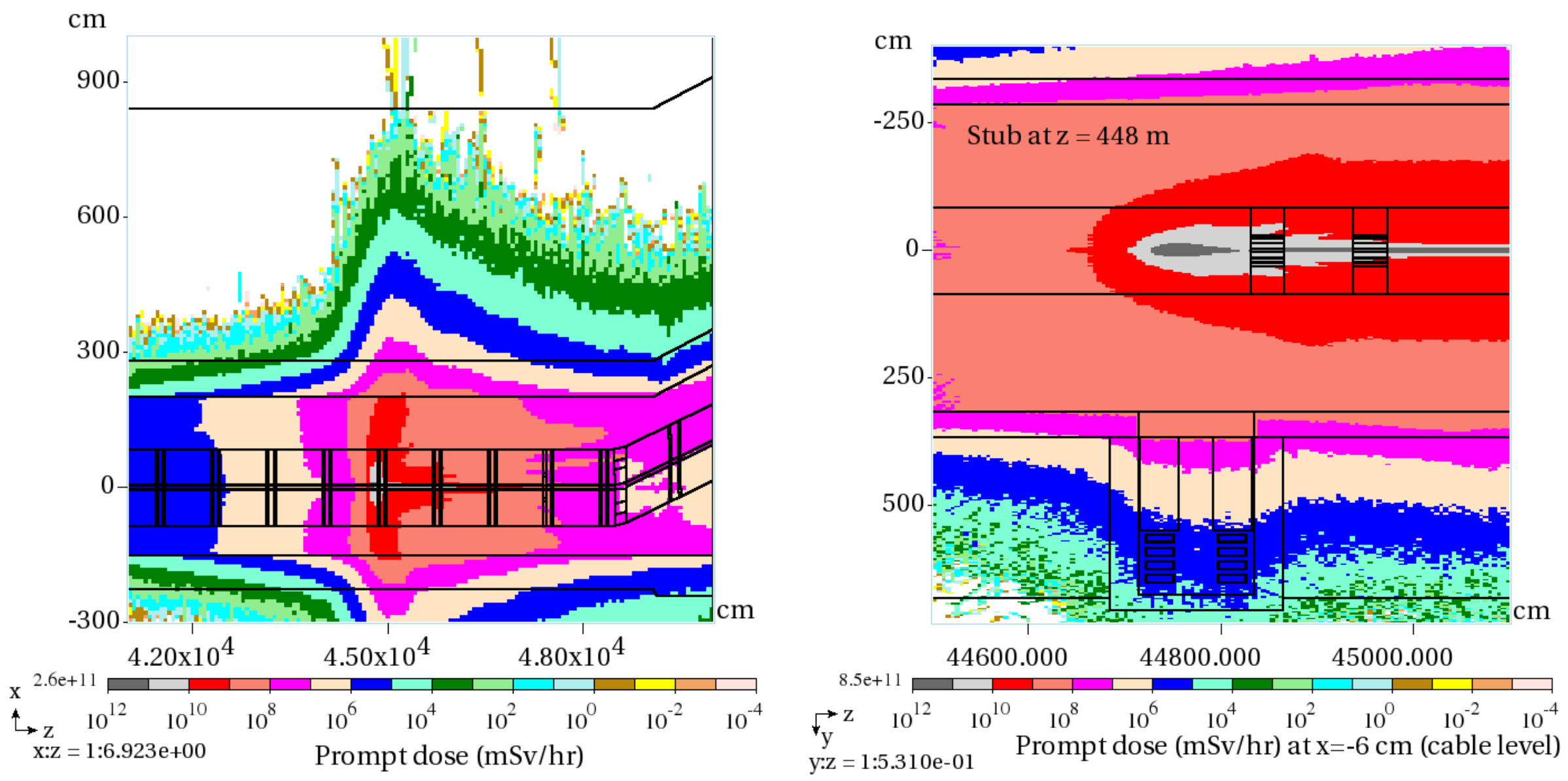

## Conclusions

Extensive MARS15 simulations were performed to determine prompt and residual dose rate maps for the European Spallation Source (ESS) linear accelerator. Both normal operations and accidental point beam loss were studied. Weak points in the accelerator shielding, such as personnel and equipment penetrations were modeled and studied carefully. Both prompt and residual dose rate values are used as input for accelerator radiological event and hazard analysis, through which required safety functions and administrative measures are defined.

## Acknowledgements

This work is supported by Fermi Research Alliance, LLC under contract DE-AC02-07CH11359 with the U.S. Department of Energy.